\documentclass[a4paper,11pt]{article}
\usepackage{pos}

\usepackage{natbib}
\title{ The X-$\gamma$ detector onboard the POEMMA-Balloon with Radio payload }

\author*[a]{Matteo Battisti}
\onbehalf{for the JEM-EUSO Collaboration\\[-1mm]{\normalsize \normalfont (a complete list of authors can be found at the end of the proceedings)}}

\affiliation[a]{Istituto Nazionale di Fisica Nucleare (INFN), Sezione di Roma Tor Vergata, Rome, Italy\\
}

\emailAdd{mbattist@roma2.infn.it}

\abstract{The POEMMA-Balloon with Radio (PBR) is a NASA mission designed to study Ultra-High-Energy Cosmic Rays (UHECRs) and Very-High-Energy Neutrinos (VHENs) from a balloon platform. Serving as a precursor to the planned POEMMA (Probe of Extreme Multi-Messenger Astrophysics) satellite mission, PBR will be launched aboard a NASA Super Pressure Balloon for a targeted flight of more than 20 days 
at an altitude of 33~km. The launch is planned for Spring 2027 from Wanaka, New Zealand. The unique conditions of low pressure and high altitude will enable in-situ observations of High-Altitude Horizontal Air Showers (HAHAs), a poorly understood class of nearly horizontal Extensive Air Showers (EASs) induced by cosmic rays skimming the Earth's atmosphere without reaching the ground. Due to the lower atmospheric grammage at these altitudes, HAHAs develop more gradually compared to typical downward-going EASs, with interaction lengths on the order of 100~km. This slow development allows balloon-borne instruments to probe the early stages of cosmic ray shower evolution.
At these early stages, high-energy electrons and positrons from the electromagnetic component of the shower can generate X-rays and gamma rays via synchrotron radiation. The X-$\gamma$ detector onboard PBR is designed to measure these photons across a broad energy range. The instrument consists of four sub-detectors, each optimized for different overlapping energy bands: X-ray (10–30 keV), X-$\gamma$ (30–300 keV), and $\gamma$-ray (100–4000 keV, with two detectors). The current design utilizes CsI(Tl)/NaI(Tl) scintillating crystals coupled with Silicon Photomultipliers (SiPMs) for photon detection. To suppress background noise, all detectors—except for the X-ray entrance window—are enclosed within an anti-coincidence system composed of plastic scintillators, also read by SiPMs, to reject charged particle events.
The X-$\gamma$ detector is aligned with PBR’s primary instruments—the Fluorescence Camera and the Cherenkov Camera—within a 30$^\circ$ field of view, overlapping with both imaging cameras. It will operate in a triggered mode, with the possibility to receive signals from the other instruments to check for simultaneous events.
This contribution will summarize the scientific objectives of the X-$\gamma$ detector and provide an overview of its design, functionality, and current development status.}

\ConferenceLogo{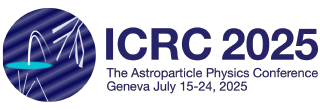}

\FullConference{39th International Cosmic Ray Conference (ICRC2025)\\
 15–24 July 2025\\
Geneva, Switzerland\\}

\begin{document}
\maketitle

\section{The PBR payload}
\begin{figure}
   \centering
   \includegraphics[width=\columnwidth]{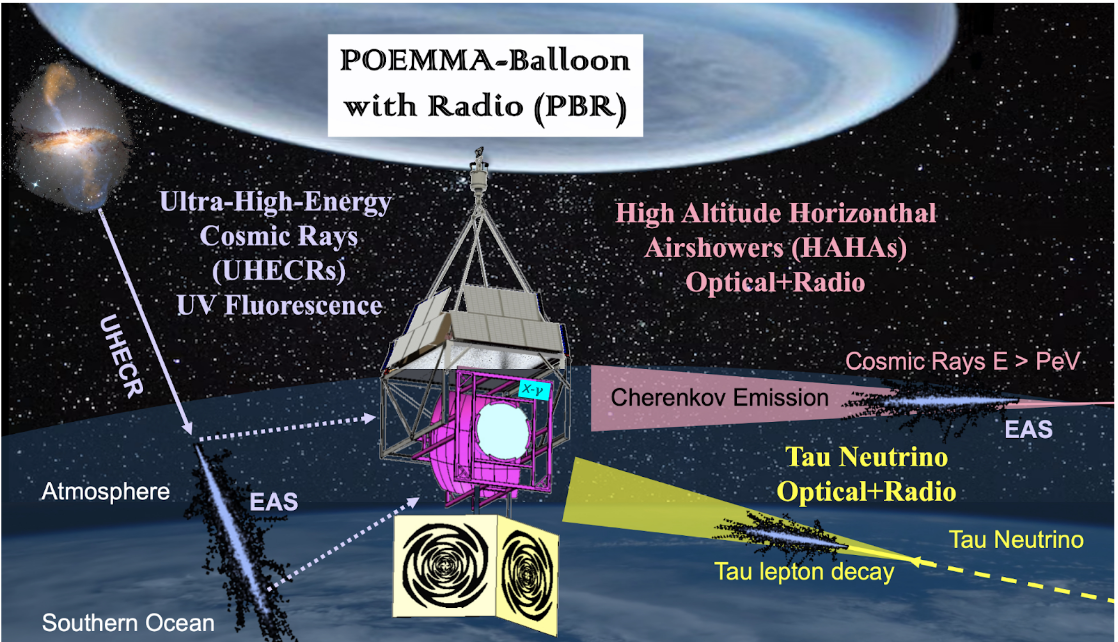}
   \caption{PBR sketch along its scientific goals. The X-$\gamma$ detector (light blue box, not in scale) will point in the same direction of the other telescopes (Cherenkov and Fluorescence cameras).}
	 \label{fig:science_sketch}
\end{figure}

Fig.~\ref{fig:science_sketch} shows a sketch of the POEMMA-Balloon with Radio (PBR) \cite{PBR_overview} payload, scheduled for launch aboard a NASA Super Pressure Balloon (SPB) from Wanaka, New Zealand in the Southern Hemisphere spring of 2027. Built on the previous experience of EUSO-SPB2 \cite{SPB2_lesson_learned}, PBR will operate at an altitude of approximately 33 km for a flight duration of more than 20 days. From this unique position, PBR will study Ultra-High Energy Cosmic Rays (UHECRs), Very-High Energy neutrinos (VHENs), and transient astrophysical events.

The PBR payload features a mechanically steerable platform capable of full 360$^\circ$ azimuthal rotation and zenith tilting from nadir to above the horizon \cite{PBR_tilt}, providing accurate pointing and tracking capabilities. At the heart of the instrument are two primary cameras: a Fluorescence Camera (FC) \cite{PBR_FC} and a Cherenkov Camera (CC) \cite{PBR_CC}, both sharing a Schmidt-based optical system with a 1.1~m diameter entrance pupil \cite{PBR_opt_and_mech}. The FC is optimized for the detection of UHECR-induced extensive air showers (EAS) with energies exceeding 2$\times$10$^{18}$~eV when pointed toward nadir. Its focal plane is populated with Multi-Anode Photomultiplier Tubes (MAPMTs), offering high spatial ($\sim$100~m on ground) and time (1~$\mu$s) resolution for the detection of UV light produced in the atmosphere by air showers.

Complementing the FC, the CC is equipped with a silicon photomultiplier (SiPM) array and is specialized for detecting fast, beamed Cherenkov flashes. It is optimized for two main science targets: (1) the detection of UHE neutrinos from transient Target-of-Opportunity (ToO) events, such as those identified by multimessenger observatories, by observing just below the limb; and (2) the study of High-Altitude Horizontal Air showers (HAHAs) generated by cosmic rays interacting in a rarefied atmosphere and observable slightly above the limb (see Sec.\ref{Sec:HAHAs})

In addition to the optical instruments, PBR includes a Radio Instrument (RI) designed to provide complementary measurements of impulsive radio signals emission of the the EAS, enhancing the multimodal detection capability. 

Finally, a X-$\gamma$ ray detector completes the payload (see Sec.\ref{Sec:X-gamma_detector}), offering a possible novel approach to the study of HAHAs and the development of EASs.

\section{X-$\gamma$ emissions from HAHAs}
\label{Sec:HAHAs}

\begin{figure}
   \centering
   \includegraphics[width=\columnwidth]{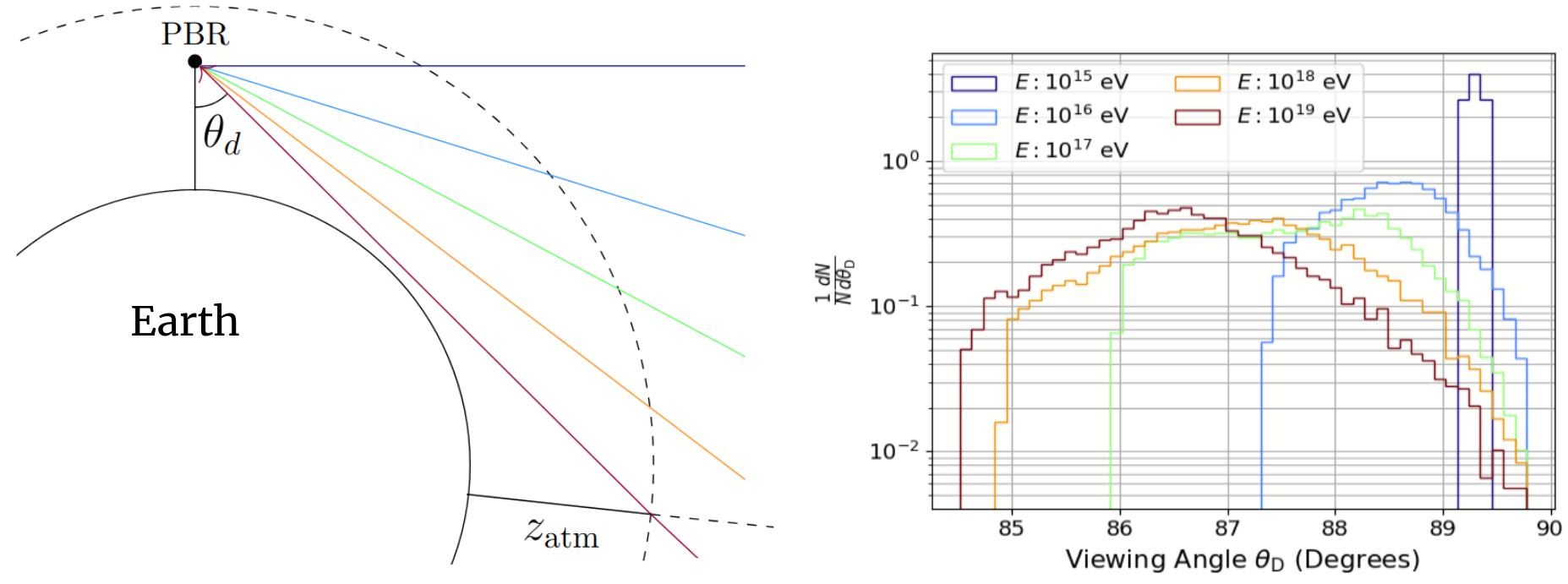}
   \caption{\textbf{Left:} Possible trajectories of HAHAs that end up reaching PBR. Picture not on scale. \textbf{Right:} . Energies of HAHAs that can reach the detector as a function of their zenith angle. 90$^\circ$ are horizontal events. Picture adapted from \cite{Cherenkov_emission}. }
	 \label{fig:HAHAs_science_1}
\end{figure}

HAHAs are EASs induced by cosmic rays that skim the Earth’s atmosphere and traverse the telescope field of view, never intersecting the ground. PBR can observe HAHAs in tilted mode looking at or above the limb (Fig.~\ref{fig:HAHAs_science_1}, left). 
In this configuration, the atmosphere itself acts as an energy filter (Fig.~\ref{fig:HAHAs_science_1}, right) since more inclined events have to cross a significantly larger portion of the atmosphere before reaching the detector. PBR will observe HAHAs ranging from Earth’s horizon (84.2$^\circ$, above 10$^{19}$~eV) to horizontal (90$^\circ$, above  10$^{15}$~eV). EASs with energy above 10$^{17}$~eV are expected above 86$^\circ$, at a rate $\sim$1 event per night of observation.
 
\begin{figure}
   \centering
   \includegraphics[width=\columnwidth]{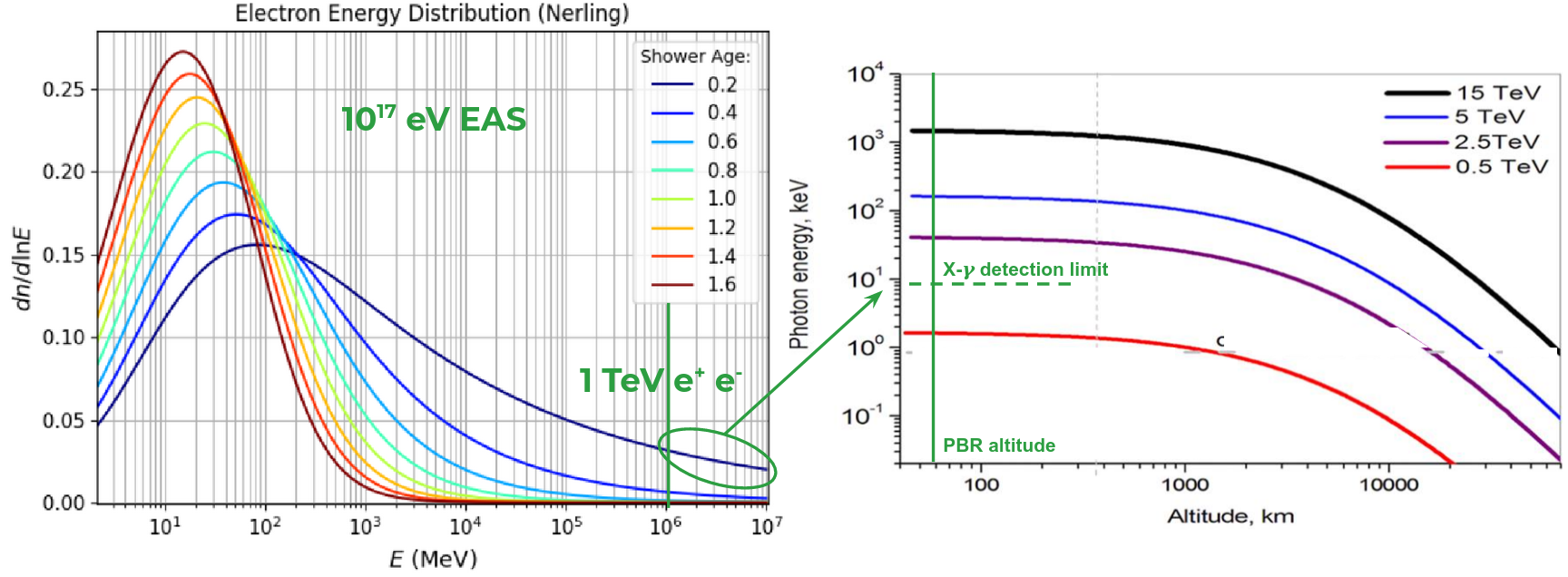}
   \caption{\textbf{Left:} Electron energy distribution for different shower ages, according to \cite{NERLING2006421} for a 10$^{17}$~eV primary. For shower ages $\lesssim$0.2 there is a relevant fraction of electrons and positrons above 1 TeV. \textbf{Right:} Average energy of synchrotron photons emitted by electrons with different energies. Adapted from \cite{Galper_2017}. Electrons and positrons at 1 TeV should produce $\sim$10 keV photons via bremsstrahlung}
	 \label{fig:HAHAs_science_2}
\end{figure}

Due to the reduced atmospheric grammage at high altitudes, HAHAs exhibit a more gradual longitudinal development compared to conventional downward-propagating extensive air showers, with characteristic interaction lengths on the order of $\sim$100~km. This extended development permits balloon-borne detectors to access the initial stages of shower evolution. The early shower development is a topic still not well understood. According to Nerling model \cite{NERLING2006421}, for a primary above 10$^{17}$~eV in the early age of the shower, there is a significant fraction of electrons and positrons above 1 TeV (Fig.~\ref{fig:HAHAs_science_2}, left). Those particles emit X-rays above 10 keV (Fig.~\ref{fig:HAHAs_science_2}, right). The long interaction length ($\sim$100 km at balloon flight altitude) allows the in-situ detection of such signals.

\section{The X-$\gamma$ instrument}
\label{Sec:X-gamma_detector}

\begin{figure}
   \centering
   \includegraphics[width=.25\columnwidth]{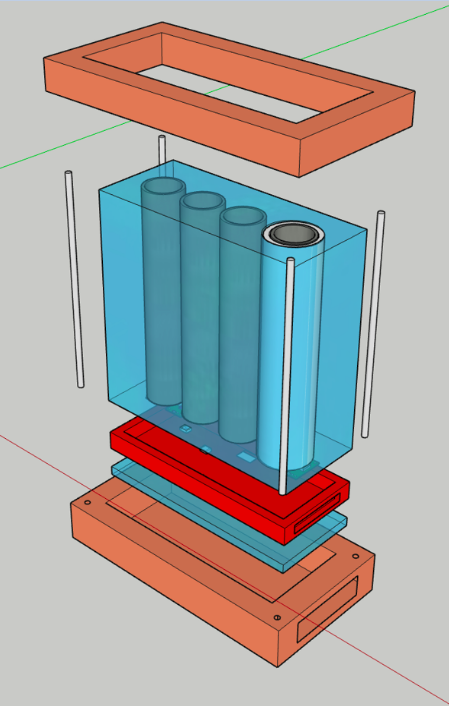}
   \includegraphics[width=.55\columnwidth]{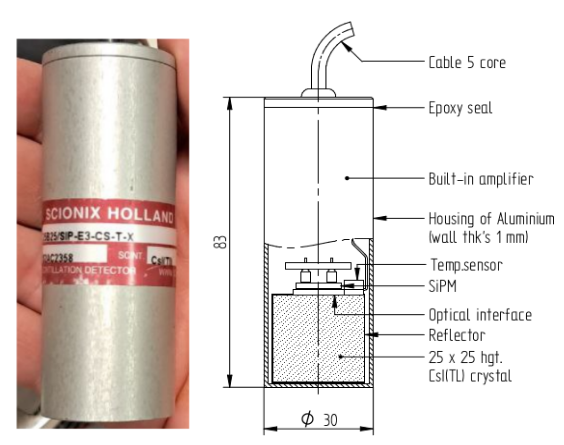}
   \caption{\textbf{Left:} Exploded view of the detector. The entire instrument fits in a $\sim$30$\times$15$\times$30~cm$^3$ box. \textbf{Right:}  Scionix detector for the X-$\gamma$ channel with its schematic. } 
	 \label{fig:detector}
\end{figure}

The PBR  X-$\gamma$ instrument is designed to observe high-energy photons starting from $\sim$10 keV. Fig.~\ref{fig:detector}, left, shows the exploded view of the entire detector. 
The primary detectors are 4 Scionix devices\footnote{https://scionix.nl/} made of scintillating crystals (NaI(Tl) or CsI(Tl)) coupled with a SiPM readout and built-in preamplifier (Fig.~\ref{fig:detector}). The devices come from the manufacturer with a built-in bias generator and temperature sensor in feedback loop with the SiPM. Three different devices are used in the PBR  X-$\gamma$ instrument, each one optimized for a different energy range: an \textbf{X Channel} [10-30 keV], an \textbf{X-$\gamma$ Channel} [30-300 keV] and two \textbf{$\gamma$ Channels} [100-4000 keV].

The X channel detector features a beryllium window of 0.3~mm thickness to reduce the X-ray absorption, while the other channels employ a 0.1~mm aluminium entrance. As shown in the left image, the 4 devices are placed inside a collimator to restrict the field of view to 16$^\circ$ for the X channel and 30 $^\circ$ for the others. Moreover, the whole detector is enclosed in a plastic scintillator structure (light blue in the picture), read-out by 4 SiPMs, used as a veto for charged particles. The entrance of the X channel is not enclosed in the veto system, again to reduce the material in front of the low-energy detector. Instead, a permanent magnet that deflects the charged particle is housed inside the collimator.

One single board with an embedded FPGA is used to power the detector, set the required parameters, manage the trigger logic and read out and digitize the 8 channels (4 from the main detectors, 4 from the veto SiPMs).

\section{Data acquisition and trigger system}
\begin{figure}
   \centering
   \includegraphics[width=.85\columnwidth]{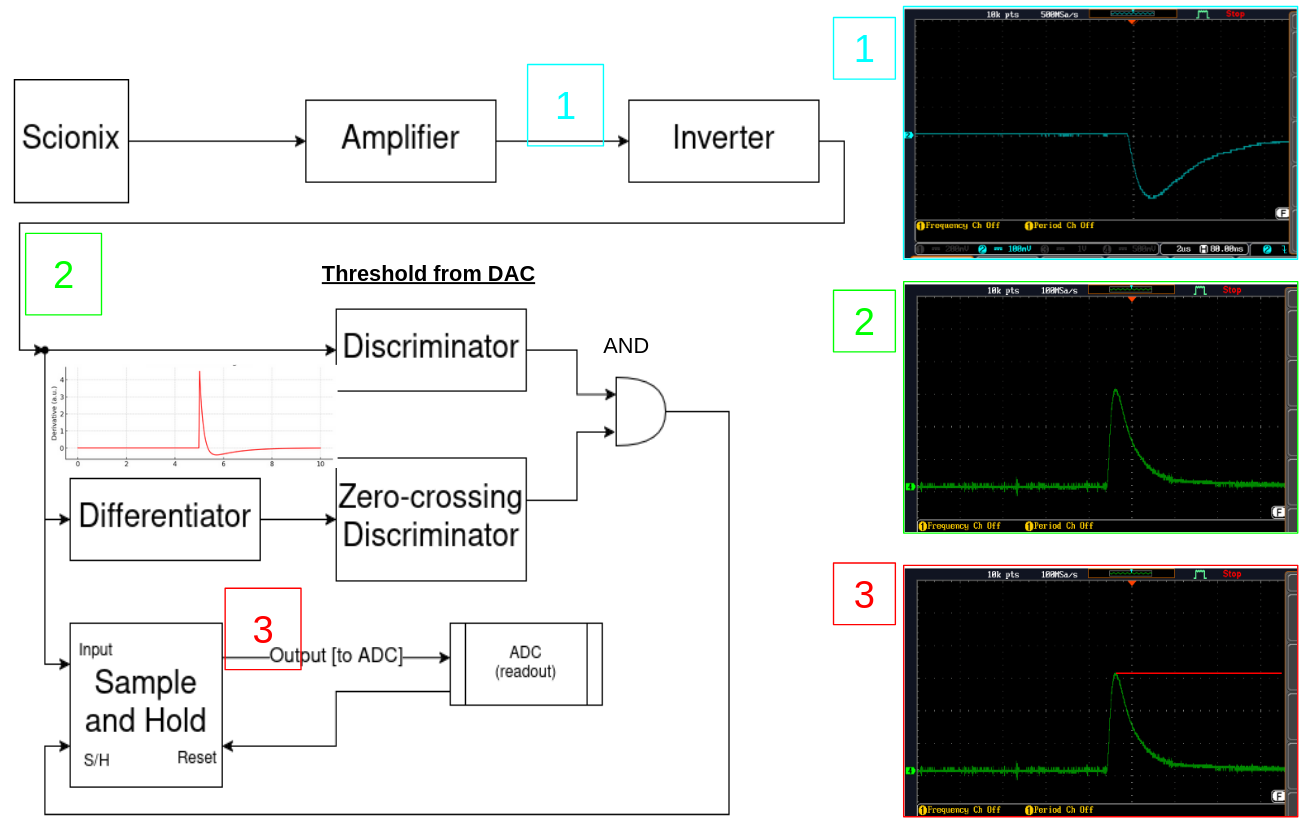}
   \caption{Simplified schematic of a single channel trigger logic. The figures on the right are oscilloscope measurements taken at the points of the electronic chain indicated by the numbers.}
	 \label{fig:trigger_logic}
\end{figure}

Fig.~\ref{fig:trigger_logic} shows the electronic chain of the signal of a single Scionix device. With reference to the numbers in the picture, the Scionix signal detector is amplified [1], inverted [2], and split into three branches. The first branch goes to a discriminator, whose threshold is set by the FPGA. A second copy of the signal is differentiated [see insert in the left figure] and analyzed by a zero-crossing discriminator to identify the peak. The AND of the two signals (positive when the signal is over threshold AND at the maximum of the peak) is used to “drive” the Sample and Hold [S\&H], whose input is the third copy of the signal. When the S\&H receives a positive signal from the AND logic port, it “holds” the value of the maximum of the input signal [3] indefinitely, until a reset command is issued. The output of the S\&H is converted by a 12-bit ADC. After the read out ($\sim$20~$\mu$s for 4 channels), the system resets and becomes ready to trigger again.

After each ADC readout, a data packet is generated containing metadata such as the event timestamp, total active and dead time, and an event ID. This packet is then transmitted to the Processing Unit \cite{PBR_DAQ}. Simultaneously, the system receives external trigger signals from the Cherenkov Camera (CC), which are stored in a dedicated packet to enable offline searches for coincident events.

The detection of a simultaneous event by both the CC and the X-$\gamma$ detector would provide strong evidence for the observation of a HAHA, offering valuable insights into the development of these poorly understood air showers. Furthermore, such coincidences can help constrain theoretical models of the early stages of shower evolution.

\section{Expected performance}
\begin{figure}
   \centering
   \includegraphics[width=.66\columnwidth]{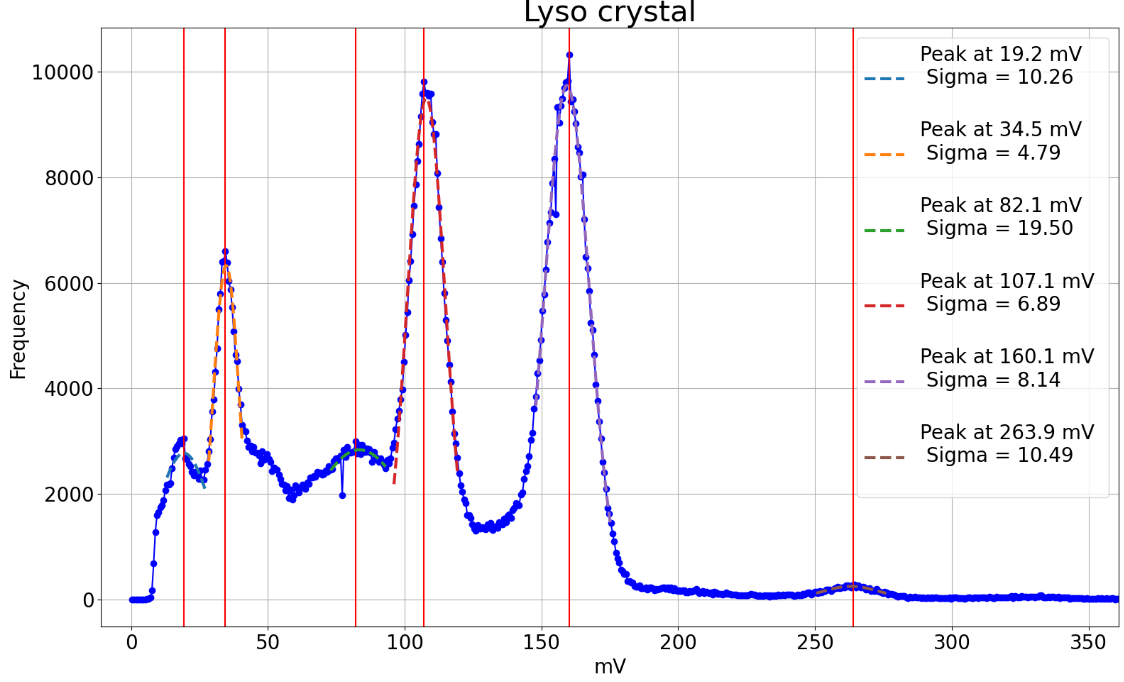}
   \includegraphics[width=.33\columnwidth]{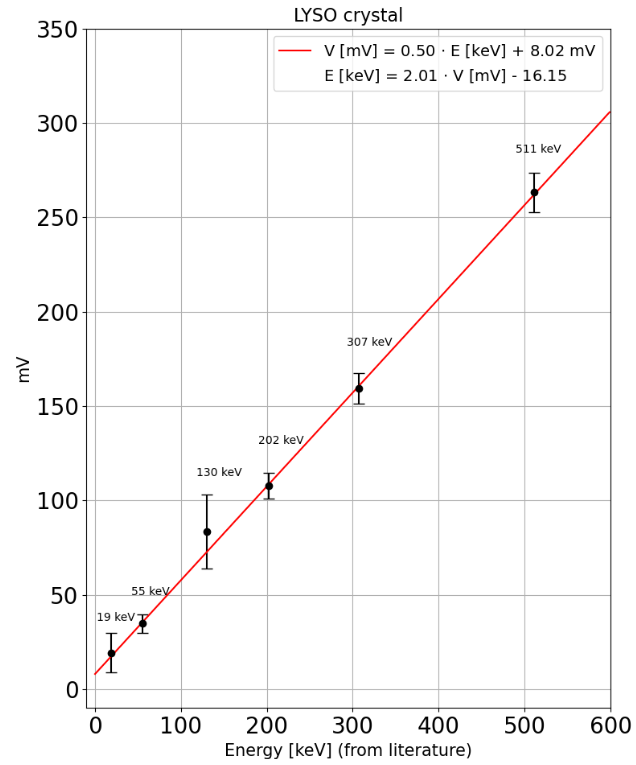}
   \caption{\textbf{Left:} spectrum of a lutetium-containing Lyso crystal measured by the X-channel. \textbf{Right:} Linearity of the detector’s response. The values on the X-axis come from the literature. }
	 \label{fig:LYSO}
\end{figure}

A scaled-down version of the X-$\gamma$ instrument has been extensively tested in the laboratory. This version is made of a single Scionix device (the X Channel) and a first version of the electronic board. Several sources containing different radioactive materials have been tested, producing plots similar to the one shown in Fig.~\ref{fig:LYSO}. This one in particular has been obtained using a LYSO (Lutetium–yttrium oxyorthosilicate) crystal as a source. The decay of lutetium in the LYSO crystal produces characteristic energy peaks at different energies. The left plot shows the detector response to the different peaks, along with their gaussian fits. The right panel shows the correlation between the energy of the peaks (obtained from the literature) and the output of the system. The system presents a linear response with no appreciable deviation from linearity in the entire energy range. The energy threshold appears to be below 20 keV, most likely between 10 and 15 keV. Similar tests will soon be conducted on the other Scionix devices.

\section*{Acknowledgements}
The authors would like to acknowledge the support by NASA award 80NSSC22K1488 and
80NSSC24K1780, by the French space agency CNES and the Italian Space agency ASI. The work is supported by
OP JAC financed by ESIF and the MEYS CZ02.01.01/00/22\_008/0004596. This research used resources of the
National Energy Research Scientific Computing Center (NERSC), a U.S. Department of Energy Office of Science User
Facility operated under Contract No. DE-AC02-05CH11231. We also acknowledge the invaluable contributions of the
administrative and technical staffs at our home institutions.

\bibliography{my-bib-database}

 \newpage
{\Large\bf Full Authors list: The JEM-EUSO Collaboration}

\begin{sloppypar}
{\small \noindent
M.~Abdullahi$^{ep,er}$              
M.~Abrate$^{ek,el}$,                
J.H.~Adams Jr.$^{ld}$,              
D.~Allard$^{cb}$,                   
P.~Alldredge$^{ld}$,                
R.~Aloisio$^{ep,er}$,               
R.~Ammendola$^{ei}$,                
A.~Anastasio$^{ef}$,                
L.~Anchordoqui$^{le}$,              
V.~Andreoli$^{ek,el}$
% ...
}
\end{sloppypar}
\vspace*{.3cm}

{ \footnotesize
\noindent
%
% ...
$^{cb}$ Universit\'e de Paris, AstroParticule et Cosmologie (CNRS), Paris, France\\
% ...
$^{ef}$ Istituto Nazionale di Fisica Nucleare (INFN), Sezione di Napoli, Naples, Italy\\
% ...
$^{ei}$ Istituto Nazionale di Fisica Nucleare (INFN), Sezione di Roma Tor Vergata, Rome, Italy\\
% ...
$^{ek}$ Istituto Nazionale di Fisica Nucleare (INFN), Sezione di Torino, Turin, Italy\\
$^{el}$ Universit\`a di Torino, Dipartimento di Fisica, Turin, Italy\\
% ...
$^{ep}$ Gran Sasso Science Institute (GSSI), L’Aquila, Italy\\
$^{er}$ Istituto Nazionale di Fisica Nucleare (INFN), Laboratori Nazionali del Gran Sasso, Assergi, Italy\\
% ...
$^{ld}$ University of Alabama in Huntsville, Huntsville, AL, USA\\
$^{le}$ City University of New York (CUNY), Lehman College, Bronx, NY, USA\\
% ...
}

\end{document}